\documentclass[twocolumn,pr,showpacs,amsmath,longbibliography,superscript,
superscriptaddress]{revtex4-1}
\usepackage{graphicx,bm}
\usepackage{epsfig,psfrag}
\usepackage{amsmath}
\usepackage{amssymb}
\usepackage{amsbsy}
\usepackage{amsthm}
\usepackage{amsfonts}
\usepackage{wasysym}
\usepackage{bbm}
\usepackage{tabularx}
\usepackage{euscript}
\usepackage{color}
\usepackage{enumerate}
\usepackage{amsfonts}
\usepackage{exscale}
\usepackage{bbold}
\usepackage{float}
\usepackage[colorlinks,citecolor=blue]{hyperref}

\def\bea{\begin{eqnarray}}
\def\eea{\end{eqnarray}}

\def\avg#1{\left\langle#1\right\rangle}

\def\---{---\hspace*{-2mm}---}

\begin{document}
\title{Room-temperature magnetism on the zigzag edges of
phosphorene nanoribbons}
\author{Guang Yang}
\affiliation{Department of Physics, Beijing Normal University,
Beijing 100875, China}
\affiliation{Beijing Computational Science Research Center,
Beijing 100193, China}
\author{Shenglong Xu}
\affiliation{Department of Physics, University of California, San Diego,
California 92037, USA}
\author{Wei Zhang}
\affiliation{Department of Physics, Beijing Normal University,
Beijing 100875, China}
\author{Tianxing Ma}
\email{Corresponding author: txma@bnu.edu.cn}
\affiliation{Department of Physics, Beijing Normal University, Beijing 100875, China}
\affiliation{Department of Physics, University of California, San Diego, California 92037, USA}
\author{Congjun Wu}
\affiliation{Department of Physics, University of California, San Diego, California 92037, USA}

\begin{abstract}
Searching for room-temperature ferromagnetic semiconductors has evolved
into a broad field of material science and spintronics for decades,
nevertheless, these novel states remain rare.
Phosphorene, a monolayer black phosphorus with a puckered honeycomb lattice
structure possessing a finite band gap and high carrier mobility, has been
synthesized recently.
Here we show, by means of two different large-scale quantum Monte Carlo
methods, that relatively weak interactions can lead to remarkable edge
magnetism in the phosphorene nanoribbons.
The ground state constrained path quantum Monte Carlo simulations reveal
strong ferromagnetic correlations along the zigzag edges, and the
finite temperature determinant quantum Monte Carlo calculations show
a high Curie temperature up to room temperature.
\end{abstract}
\pacs{ 75.50.Pp, 81.05.Zx, 75.10.Lp, 85.75.-d}

\maketitle
\section{Introduction}

Semiconducting materials exhibiting high-temperature ferromagnetism play
a key role in realizing spintronics applications\cite{Koji2006}.
Nowadays, this class of novel materials continues to attract widespread
attention both theoretically and experimentally\cite{Wolf2001}.
Various properties of the targeted materials, including two dimensionality,
high Curie temperature, high carrier mobility, and the intrinsically insulating
bulk suitable for charge carrier doping\cite{Wolf2001,Ohno2000,Dietl2002,Koji2006}, are required for the high-performance nonvolatile transistors.
After intensive studies for decades, the realization of ideal ferromagnetic
semiconductors satisfying all these requirements\---\textit{in} particular, with
Curie temperatures at the scale of room temperature\---remains a challenging
problem.


Recently, two-dimensional materials, initiated by the study of graphene
and then followed by the hexagonal boron nitride and transition-metal
dichalcogenides, have attracted a great deal of attention\cite{Novoselov2004,Zhang2005}.
In graphene, the low energy physics governed by the massless Dirac
fermions leads to magnetic properties along the zigzag edges\cite{Son2006,Castro2009,Das2011,Ma2010,Feldner2011,Tao2011,Magda2014, Cheng2015,Peres2009}, which is remarkable because correlation effects in $p$-orbital materials are typically not so
strong as those in $d$- and $f$-orbital bands to drive ferromagnetism.
Experimentally, this edge ferromagnetism has been observed
\cite{Magda2014}.
The key is the appearance of the edge flat band, and the
divergence of density of states amplifies the interaction effects.
Ferromagnetism has also been investigated in the $p_x(p_y)$-orbital bands due to either the bulk flat band
structure in the honeycomb lattice\cite{wu2007,zhangsz2010}, or the quasi-one-dimensional (quasi-1D) bands in the square and cubic lattices\cite{li2014,xu2015}.
However, the zero band gap of graphene limits its application performance
as a semiconductor\cite{Tatiana2001,Wu2007,Miller2000,Fern2007,Yazyev2008,Yan2009,Zhang2005}.

It is highly desired to discover a two-dimensional material exhibiting a
finite band gap as the basis for a low-power transistor.
For this purpose, phosphorene, a puckered honeycomb structure of
monolayer black phosphorus held together by van der Waals forces,
has been isolated recently, which further advances the
development of the post-graphene materials.
Numerous first-principle studies have appeared on phosphorene\cite{Li2014, Liu2014, Xia2014,Gomez2014,Koenig2014,Peng2014,Tran2014,Qiao2014,Fei2014,Rudenko2014,Ezawa2014,Sisakht2014}
and its nanoribbons\cite{Tran2014B,Guo2014,Carvalho2014,Peng2014JAP,Zhu2014,Du2015}.
Different from the graphene structure, as shown in Fig.\,\ref{Fig:Sketch}(a), the hopping integrals in phosphorene are strongly anisotropic.
As a result, its electronic spectrum changes significantly
from that of graphene: Phosphorene is fully gapped as a direct gap semiconductor\cite{Du2015}.
Furthermore, its mobility is higher compared with those of transition-metal
dichalcogenides\cite{Li2014}. Therefore, phosphorene is considered as a promising candidate for
future spintronics applications\cite{Peng2014,Rudenko2014,Ezawa2014,
Sisakht2014}. It is natural and important to further investigate
the possible magnetic properties in phosphorene\cite{Du2015}.


It has been theoretically proposed 
that applying strain along the zigzag edges can enhance the graphene
edge ferromagnetism\cite{Peres2009,Cheng2015}.
The zero energy edge modes only occupy part of the one-dimensional edge
Brillouin zone connecting the projections of two bulk Dirac points.
The stress induced anisotropy shifts the locations of the Dirac points,
and thus modifies the interaction effect by changing
the density of states in the flat band along the edges.
Nevertheless, the available strain strengths are limited in graphene\cite{Peres2009},
and applying strain significantly degrades mobility\cite{Tang2015}.
In contrast, the intrinsic anisotropy in phosphorene is about one order
larger than most severely strained graphene\cite{Peres2009}.
Consequently, the phosphorene nanoribbon exhibits a quasiflat edge
band across the entire one-dimensional edge Brillouin zone, which
is entirely detached from the bulk band\cite{Carvalho2014,Peng2014JAP}.
We anticipate much more enhanced edge ferromagnetism in phosphorene
than in graphene.
Regarding the natural strong anisotropy, the great carrier
mobility, and the direct gap, the high-temperature ferromagnetism
in phosphorene nanoribbons is not only of great academic interest
from the aspect of the interplay between band structure and
interaction in solids, but also paves the way for novel
technology revolutions. 

In this paper, we employ the nonperturbative numeric methods of
large-scale quantum Monte Carlo simulations to investigate
the edge magnetism in the bulk insulating phosphorene nanoribbons.
Strong edge ferromagnetic correlations are observed at zero temperature
with weak interactions.
The edge magnetic susceptibility at finite temperature exhibits
the Curie-Weiss behavior with the extrapolated Curie temperature
up to room temperature.
It is expected that such edge magnetism could be detected by scanning
tunneling microscopy in undoped or low doped phosphorene nanoribbons.

\section{Model and methods}
\begin{figure}[bp]
\includegraphics[width=8.5cm]{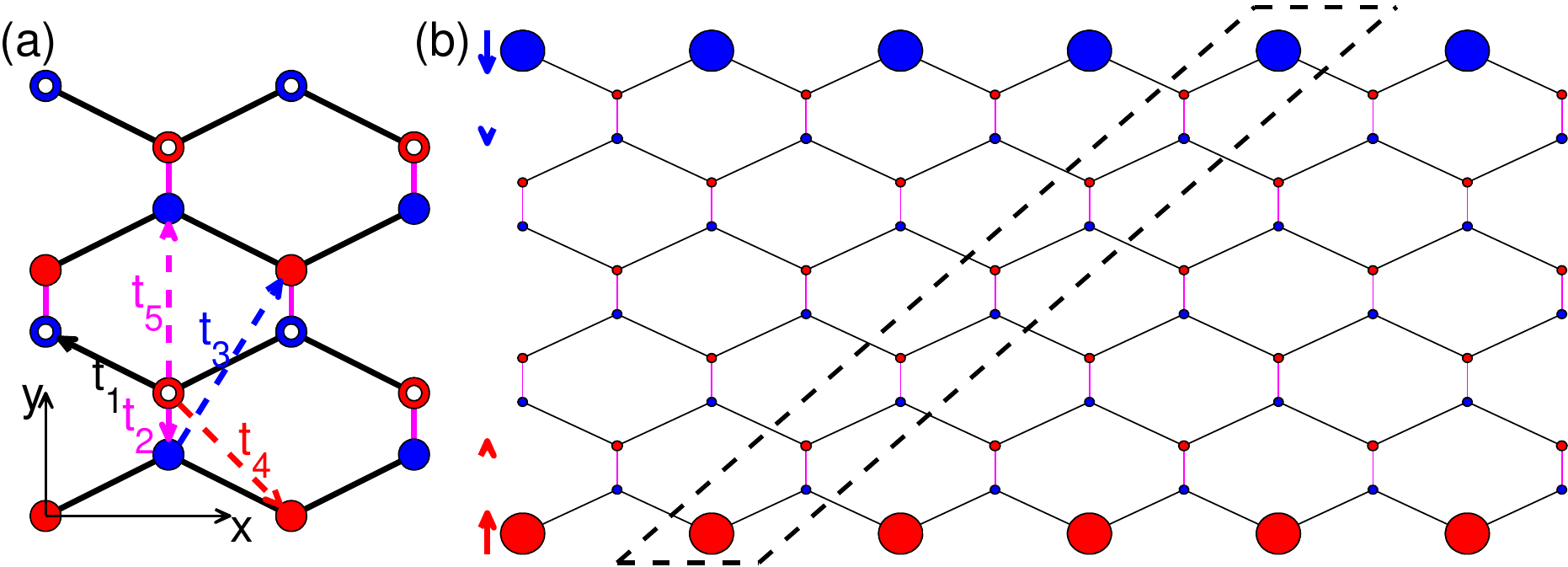}
\caption{(Color online) (a) Top view of the puckered honeycomb structure of phosphorene.
Different colors indicate atoms on different sublattices, and
solid (empty) circles indicate upper (lower) layers. The hopping parameters $t_{1}-t_{5}$ are labeled between the two corresponding sites.
(b) The schematic diagram for the  spatial distribution of the magnetic structure factor for $U=3.0$ eV at half filling on the lattice of phosphorene nanoribbon used in our simulation. The radius of each circle is proportional to the value of $M_{\bf R}$ [Eq. (\ref{Mr})]
of the row. We implement the periodic boundary condition along the $x$ direction and a unit cell is enclosed by the
dashed lines.}
\label{Fig:Sketch}
\end{figure}

To capture the physical properties of phosphorene, the tight-binding
model has been constructed recently by including five hopping integrals
$t_{i}$ among neighboring sites ($i=1,2,\ldots,5$) \cite{Rudenko2014},
as illustrated in Fig.\,\ref{Fig:Sketch}(a).
Different colors respectively represent atoms on different sublattices,
and solid (empty) circles indicate upper (lower) layers.
$t_{1,2}$ describe the two nonequivalent nearest-neighbor bonding,
$t_3$ is a next-nearest-neighbor hopping integral, and $t_{4,5}$ describe
bondings between two different next-next-nearest-neighbor hopping integrals.
These hopping integrals were fitted by the following values as $t_1=-1.220$ eV,
$t_2=3.665$ eV, $t_3=-0.205$ eV, $t_4=-0.105$ eV, and $t_5=-0.055$ eV for
bonds shown in Fig.\ref{Fig:Sketch}(a).

To construct a nanoribbon with the zigzag edges, we define the lattice shown
in Fig.\,\ref{Fig:Sketch}(b) to be periodic along the $x$ direction and finite in
the $y$ direction.
We employ the single-band Hubbard model for phosphorene nanoribbons as
\begin{equation}
H=\sum_{\langle {\bf ij} \rangle\sigma} t_{\bf ij} c^{\dag}_{{\bf i}\sigma}
c_{{\bf j}\sigma}^{\phantom{\dag}} + U\sum_{{\bf i}} n_{{\bf i}\uparrow}n_{{\bf i}\downarrow}
-\mu \sum_{\langle {\bf i} \rangle\sigma} c^{\dag}_{{\bf i}\sigma}c_{{\bf i}\sigma}^{\phantom{\dag}},
\label{eq:Ham}
\end{equation}
where the summation runs over the lattice sites.
$t_{\bf ij}$ is the hopping integral between the $ {\bf i}$ th and ${\bf j}$ th
sites, and $c^{\dag}_{{\bf i}\sigma}$ ($c_{{\bf j}\sigma}^{\phantom{\dag}}$) is the creation
(annihilation) operator of electrons at site ${\bf i}$ (${\bf j}$).
$\mu$ is the chemical potential and $U$ is the on-site repulsion.

We solve the interacting Hamiltonian equation (\ref{eq:Ham}) by using the determinant
quantum Monte Carlo (DQMC) method at finite temperature\cite{Blankenbecler1981, Hirsch1985,Raimundo2003}, and the constrained path
quantum Monte Carlo (CPQMC) for the ground-state properties\cite{Zhangcpmc}.
Both of them are nonperturbative methods suitable for simulating magnetic
correlations in the presence of Coulomb interactions\cite{Ma2013,Madqmc,Macpmc}.
The strategy of the DQMC is to represent the partition function as a
high-dimensional path integral of imaginary time evolution over a
set of random auxiliary fields,
and the integral is then performed by the stochastic importance sampling.
In the CPQMC method, the ground-state wave function is projected from an
initial trial one through a random series of Slater determiant
wave functions dependent on the auxiliary field configurations
during the imaginary time evolution.
At each step of this evolution, only the Slater determinant
wave functions with positive overlap with the initial wave function
are kept.
For more technique details we refer to Refs.~\cite{Hirsch1985,Blankenbecler1981,Raimundo2003,
Zhangcpmc,Ma2013,Madqmc,Macpmc}.

The magnetic properties are probed in several ways.
As will be shown below, only edges exhibit prominent magnetism
and the insulating bulks remain nonmagnetic.
To explore the thermodynamic properties of the edge magnetism,
we calculate the uniform magnetic susceptibility $\chi$ along each edge
by using the DQMC at finite temperatures.
The uniform spin susceptibility is the zero-frequency correlation
which is equivalent to the equal-time correlation when spins are conserved.
However, spins along the edges are not conserved by themselves;
the zero-frequency correlation is no longer equal to the equal-time
one.
The edge magnetic susceptibility $\chi$ are defined as
\begin{eqnarray}
\chi= \int_{0}^{\beta}d\tau \sum_{\bf i,j}
\langle\textrm{S}_{\bf i}(\tau) \cdot \textrm{S}_{\bf j}(0)\rangle
\label{Chi}
\end{eqnarray}
where $S_{\bf i}(\tau)=e^{H\tau}S_{\bf i}(0)e^{-H\tau}$,
and the summation is over the sites along a single edge first, and
then make an average over the results from both the top edge and the
bottom edge.
Specifically, the sites within the two edges are marked with larger
circles shown in Fig.\,\ref{Fig:Sketch}(b).
In order to further extract the spatial distribution of the magnetic
correlations, we use the CPQMC method to calculate the equal-time
magnetic structure factor for each row parallel to the zigzag boundary
defined as
\begin{eqnarray}
M_{\bf R}= \frac{1}{L^2_x}\sum_{\bf i,j\epsilon Row} S_{\bf {i,j}},
\label{Mr}
\end{eqnarray}
where $S_{\bf {i,j}}=\avg{{\bf S_i} \cdot {\bf S_j}}$,
${\bf S_i}=c^\dagger_{i}{\bf{\sigma}} c_{i}$ is the on-site spin operator,
${\bf R}$ is the index of the row, ${\bf {i,j}}$ are the site indices
along the ${\bf R}$th row, and $L_x$ is the number of sites in each row.
$M_{\bf R}$'s are calculated along each row from the bottom edge, via the center, to the top edge as shown in Fig.\,\ref{Fig:Sketch}(b),
where a schematic diagram for the spatial distribution of spin structure factor $M_{\bf R}$ is presented, in which
the radius of each circle is proportional to $M_{\bf R}$ for each row.
Strong ferromagnetic correlations appear along the edges, while those in the bulk are much weaker.
It further shows that most electrons with spin up occupy the bottom edge,
and thus most electrons with spin down occupy the top edges.

\vspace{0.15in}

\begin{figure}[t]
\includegraphics[width=0.95\columnwidth]{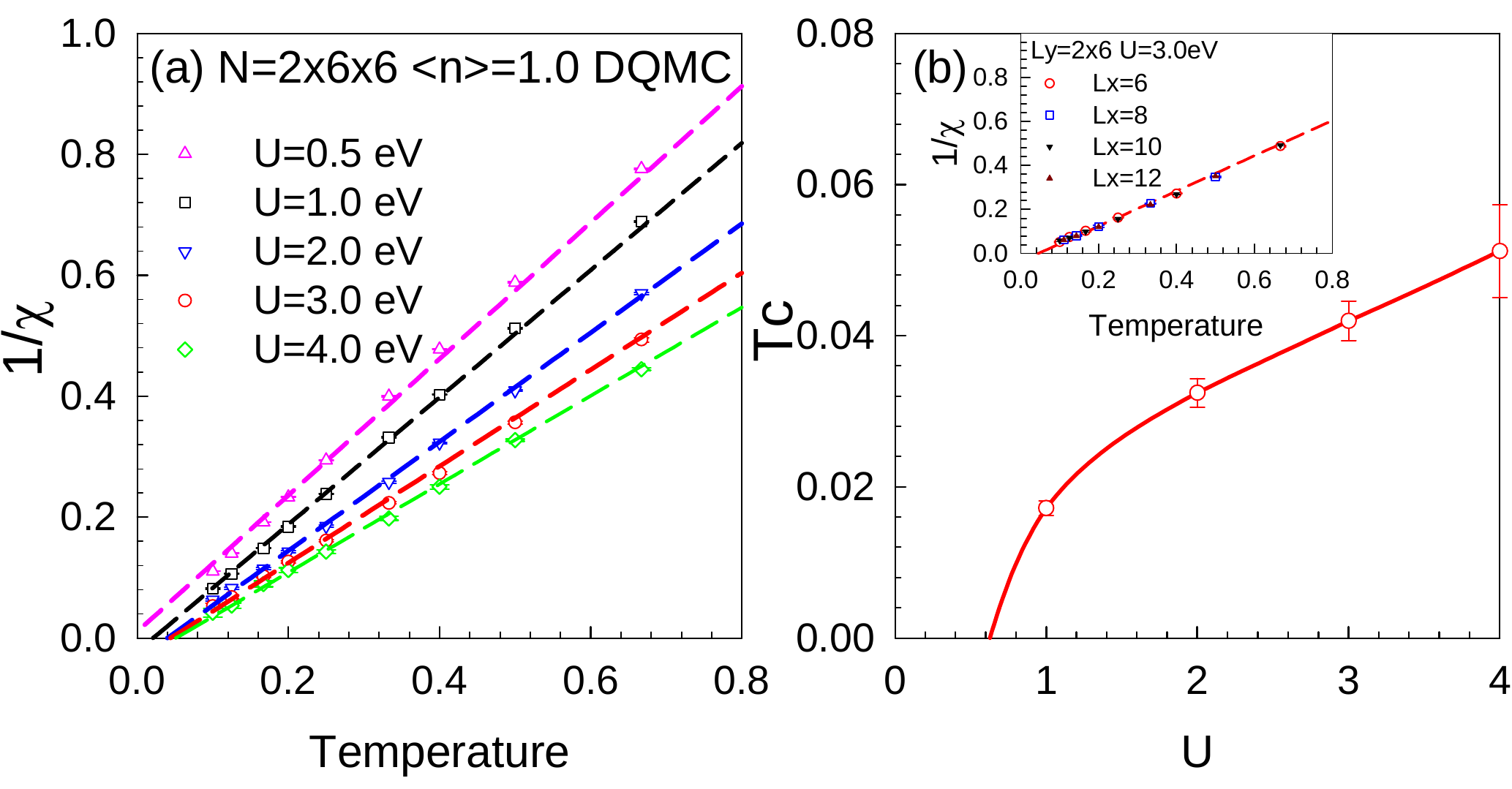}
\caption{(Color online)(a) The temperature-dependent $1/\chi$ at
$\avg{n}=1.0$ with different $U$. (b) The estimated $T_c$ depends on
Coulomb interaction $U$ at half filling. Inset: the temperature-dependent $1/\chi$ at
$\avg{n}=1.0$ with $U=3.0$ eV for different lattice size.}
\label{Fig:Xri}
\end{figure}
\section{Results and discussions}
Now we present the temperature dependence of the edge magnetic susceptibility.
In Fig.\,\ref{Fig:Xri}(a), $1/\chi(T)$ (symbols) are presented at different
interaction strengths $U$ for $\avg{n}=1.0$ as well as the linear fittings
(dashed lines).
They exhibit the Curie-Weiss behavior $1/\chi=(T-T_c)/A$.
Specially, as $U$'s are larger than 1 eV, the interceptions of the
extrapolations of $1/\chi(T)$ on the $T$ axis are finite yielding
a finite $T_c$.
Depending on the fitting, $T_c$ is estimated as $\sim$\,$0.032$ eV
for $U=2.0$ eV, which is roughly $\sim$\,$320$ K.
In Fig.\,\ref{Fig:Xri}(b), the estimated interaction dependence
of $T_c$ is shown, which indicates that $T_c$ increases as
increasing $U$. 
In the inset of Fig.\,\ref{Fig:Xri}(b), $1/\chi(T)$ (symbols) at $U=3.0$ eV are presented for $\avg{n}=1.0$ with different lattice size, as well as the linear fittings (dashed lines). Results
for 2$\times 6 \times 6$, 2$\times 8 \times 6$, 2$\times 10 \times 6$, and 2$\times 12 \times 6$ are almost the same within the
error bar. Hence we may argue here that the magnetic susceptibility and the estimated $T_c$ are almost independent of the lattice size. 

For higher interactions shown in Fig.\,\ref{Fig:Xri}(b), one may notice significant error bars on the
estimated $T_c$, related to the Monte Carlo sampling. For the susceptibility shown in Fig.\,\ref{Fig:Xri}(a) and spin correlation shown in further figures, most error bars are
controlled with 10\%. 
Where not shown in figures, error bars are within the symbol size.

\begin{figure}[t]
\includegraphics[width=0.95\columnwidth]{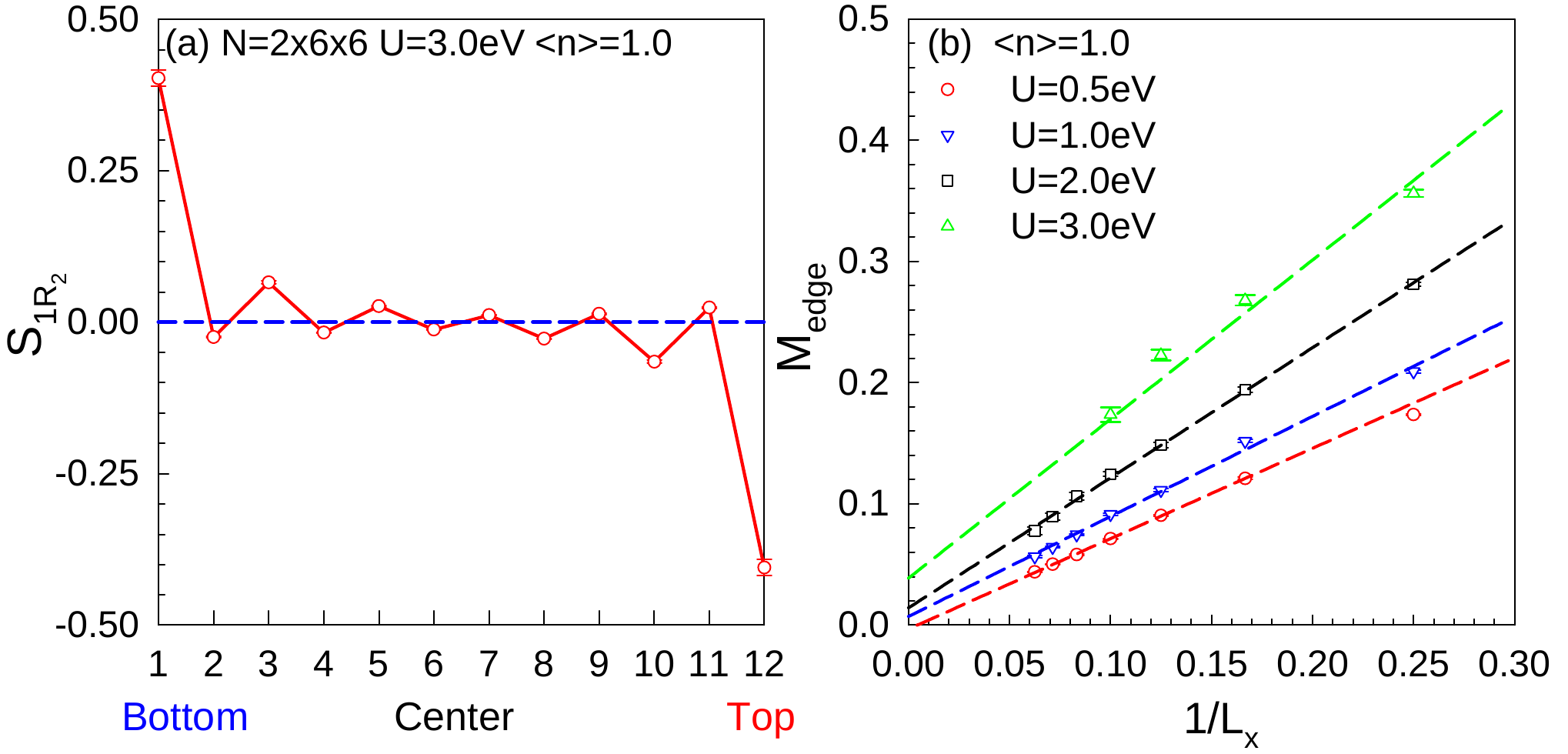}
\caption{(Color online) (a) The spin correlation
$S_{\bf {1R_2}}$ between the site $\bf 1$ and the second site of each row.
(b) The magnetic structure $M_{\bf edge}$ on the edge depends on $1/L_x$ at $\avg{n}=1.0$ for different $U$. }
\label{Fig:Mz}
\end{figure}

We further study the spatial distribution of the magnetic
correlations by using the CPQMC method.
To represent spin correlations between different rows, we plot $S_{1R_2}$
in Fig.\,\ref{Fig:Mz}(a) defined as the correlation between site
${\bf 1}$ at the first row and the second site of each row, $\bf {R_2}$.
The inter-row magnetic correlations are antiferromagnetic and
decay quickly as $\bf{R_2}$ enters the bulk.
Each edge exhibits
strong ferromagnetism, while two different edges are correlated
in an opposite way.
In Fig.\,\ref{Fig:Mz}(b), we show that the $M_{\bf R}$
on the edge depends on $L_{x}$ for a fix $L_{y}=4$.
After a careful scaling analysis, one may see that $M_{\bf edge}$ tends to have a long-range order as $U$ is larger than a $U_c$\,$\sim$\,$0.5$ eV, which agrees with the results shown in Fig.\,\ref{Fig:Xri}(b). The edge magnetism reported here is stronger than that in graphene
nanoribbons, and the critical interaction strength is lower
than that of graphene-based materials\cite{Magda2014,Peres2009,Cheng2015}.

The change of the topological structure induced by the natural strong
anisotropy in phosphorene is the key to understand the enhancement of the
edge ferromagnetism compared with the isotropic case for graphene.
Let us only focus on the nearest-neighboring hopping $t_1$ and $t_2$.
For the case of graphene $t_1=t_2$, using the translational symmetry
along the $x$ direction of the nanoribbon, the band Hamiltonian
reduces to a one-dimensional Su-Schrieffer-Heeger (SSH) model
along the $y$ direction with the open boundary condition, labeled by the momentum $k_x$,
\begin{equation}
H_{SSH}(k_x)=\tilde{t}_{o}(k_x)\sum_{i} c^\dagger_{2i-1}c_{2i}+\tilde{t}_{e}\sum_{i} c^\dagger_{2i}c_{2i+1}+H.c.
\end{equation}
where the hopping amplitude on odd bonds $\tilde{t}_{o}(k_x)=2t_1
\cos \frac{k_x}{2} $ and that on even bonds $\tilde{t}_{e}=t_2$.
It is well known that the system has two zero edge modes at
$|\tilde{t}_{even}|>|\tilde{t}_{odd}|$.
As long as $|t_2|>2|t_1|$, the condition is satisfied for all the values
of $k_x$, and the flat band extends over the entire Bouillon zone\cite{Ezawa2014},
detaching from the gapped bulk spectrum.
In contrast, at $|t_2|<2|t_1|$, there exist two Dirac points in the
spectrum where the flat band is terminated.
In particular, for the isotropic case of graphene, the length of the edge
flat band in the one-dimensional Brillouin zone is $\frac{2\pi}{3}$.
Therefore, the number of zero modes in the strong anisotropic case
studied here is tripled. Upon turning on $U$, ferromagnetism develops in these edge flat bands
due to the enhanced interaction effect by the divergence of the density
of states.
The full band gap in the phosphorene shortens the localization length
of the edge states and weakens the coupling between the edge and
bulk states.
When taking into account the intersublattice hopping $t_3$ and $t_5$,
the chiral symmetry of the band structure is still maintained,
and the edge modes remain flat at zero energy.
Nevertheless, the small intrasublattice hopping $t_4$ indeed breaks
the chiral symmetry.
The edge modes develop a weak dispersion with the bandwidth determined
by $t_4$, nevertheless, they remain a narrow band and
are still detached from the bulk spectrum.

\begin{figure}[t]
\includegraphics[width=0.72\columnwidth]{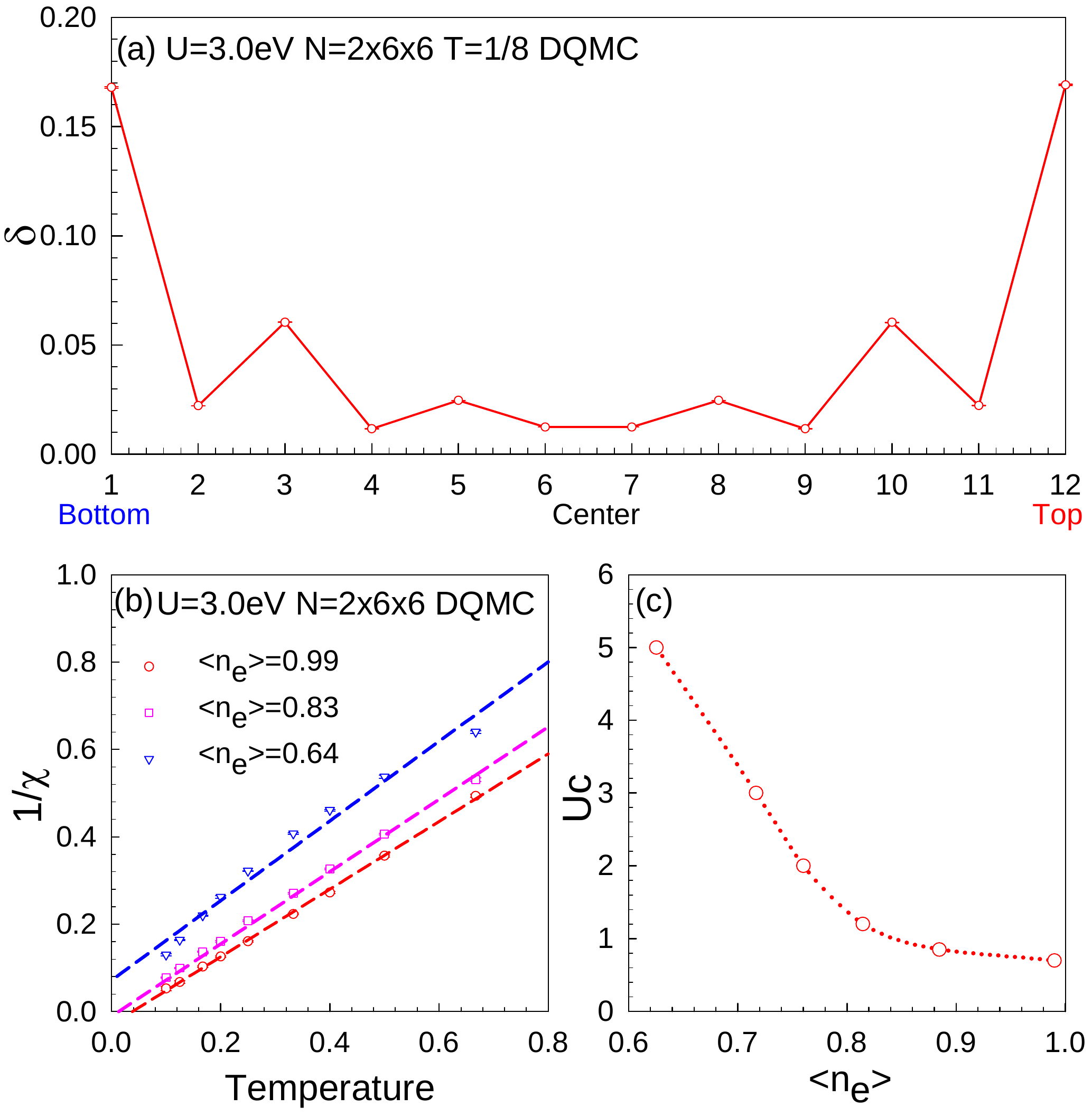}
\caption{(Color online) (a) The doped charge $\delta$ for each row parallel to the boundary. (b) The temperature-dependent $1/\chi$ with $U=3.0$ eV for different electron fillings. (c) The critical interaction $U_{c}$ as a function of electron fillings $\avg{n_{e}}$.}
\label{Fig:Uc}
\end{figure}

Next, we study the effect of doping on the edge magnetism.
In such nanoribbon with the zigzag edges, most of the doped
charge distributes along the edge, as that shown in Fig.\,\ref{Fig:Uc}(a), where the doping
$\delta=1-\avg{n_e}$, on each row parallel to the boundary is shown.
Thus in the following Figs.\,\ref{Fig:Uc}(b) and \ref{Fig:Uc}(c), we present the temperature dependence of
$1/\chi(T)$ for different edge electron fillings.
In Fig.\,\ref{Fig:Uc}(b), comparing results with different doping $\delta$ at $U=3.0$ eV, $\chi(T)$ is weakened as the system is doped away from half filling.
$1/\chi(T)$ at half filling $\avg{n_e}=0.99$ and $\avg{n_e}=0.83$ extrapolates
to a finite interception at the $T$ axis indicating a finite
Curie temperature, in contrast with $1/\chi(T)$ at  $\avg{n_e}=0.64$, in which case stronger interaction is expected to induce the edge ferromagnetism. The critical  value $U_c$ can be estimated as the smallest interaction with positive extracted $T_c$ for a fixed $\avg{n_e}$. In Fig.\,\ref{Fig:Uc}(c), we plot the phase diagram obtained this way. The critical $U_c$ is enhanced by higher doping, suggesting possible manipulation of the edge magnetism by gate voltage \cite{Novoselov2004,Zhang2005}.

\begin{figure}[t]
\includegraphics[width=1.00\columnwidth]{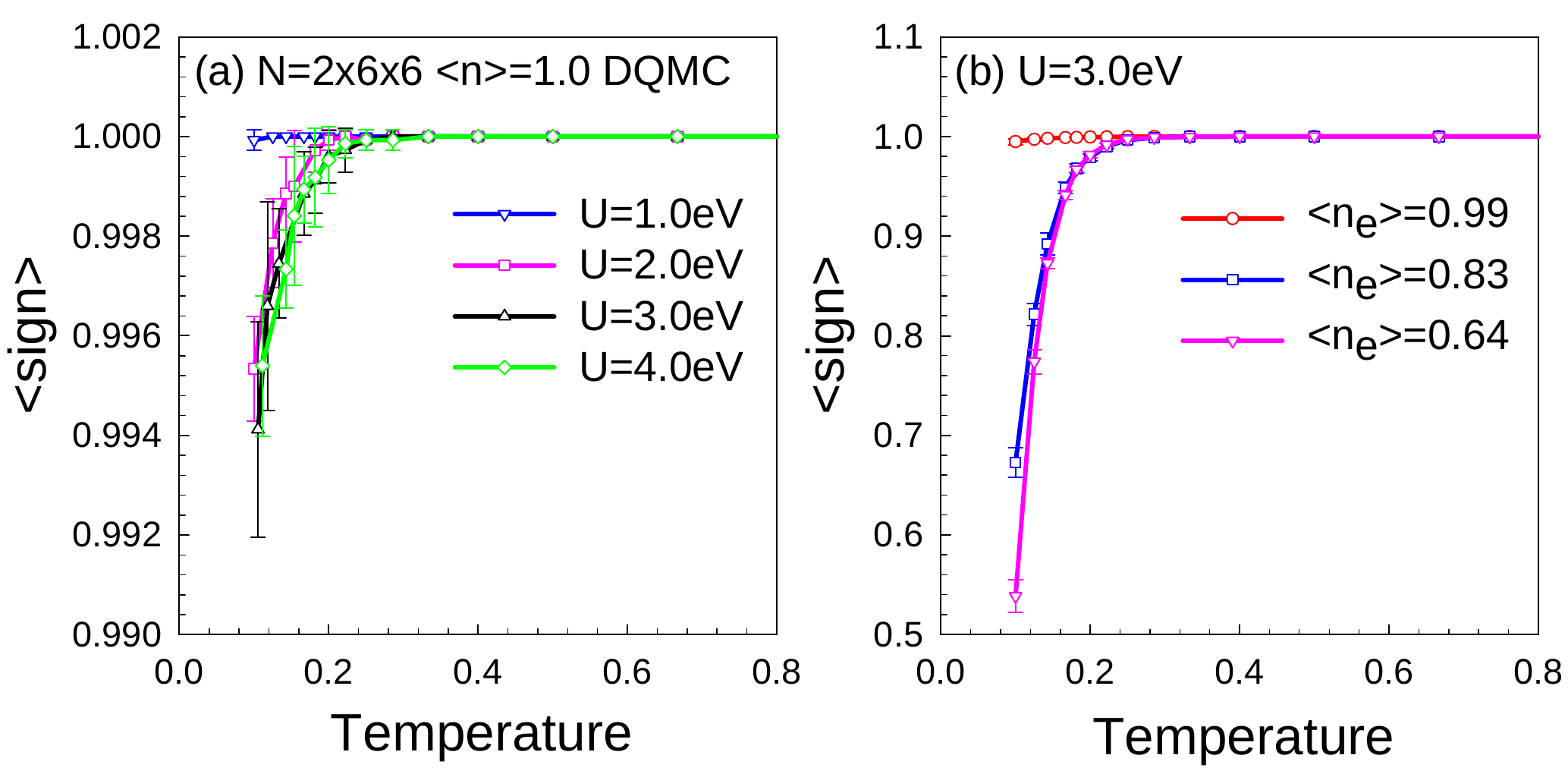}
\caption{(Color online)(a) The temperature-dependent $\langle$sign$\rangle$ at $\avg{n}=1.0$ with different $U$. (b) The temperature-dependent $\langle$sign$\rangle$ with $U=3.0$ for different electron fillings.}
\label{Fig:Sign}
\end{figure}
For the finite temperature quantum Monte Carlo method, the notorious sign problem prevents exact results for
lower temperature, higher interaction, or larger lattice. To examine the reliability of the present data, we show the average of sign in Fig. 5, dependent on different temperature $T$ at different interaction $U$ (a) and different electron fillings (b) with the Monte Carlo parameters of $30\,000$ times runs. For the results presented in previous figures at half filling, our numerical results are reliable as one can see that the average of corresponding sign is mostly larger than $0.99$ for the $U$ from 1.0 to 4.0 eV with $30\,000$ times measurements. For electron fillings away from the half filling, as that shown in Fig. 5(b),
 the average of sign decreases as the temperature is lowering, while it is larger than $0.5$ for the lowest temperature we reached. In order to obtain the same quality of data as $\langle$sign$\rangle \simeq 1$, much longer runs are necessary to compensate the fluctuations. Indeed, we can estimate that the runs need to be stretched by a factor on the order of $\langle$sign$\rangle ^{-2}$\cite{Raimundo2003}. In our simulations, some of the results are obtained with more than $120\,000$ times runs, and thus the results for the current parameters are reliable. 

To conclude, we have demonstrated that phosphorene nanoribbons, which are anisotropic direct gap semiconductors with high mobility,  exhibit strong edge magnetism with high Curie temperatures. These properties make them exceptional materials for electronic and spintronic devices. The strong filling dependence of the edge magnetism indicates flexible control on the magnetic properties of the phosphorene nanoribbons. This opens up new possibilities of engineering room-temperature electronic and spintronic devices.

\section{Acknowledgements}
This work is supported by NSFC (Grants No. 11374034 and No. 11334012) and
the Fundamental Research Funds for the Central Universities (Grant. No. 2014KJJCB26).
 We also acknowledge support from the HSCC of Beijing Normal University, and the Special Program for Applied Research on Super Computation of the NSFC-Guangdong Joint Fund (the second phase).


\end{document}